\documentclass[submission,copyright,creativecommons]{eptcs}
\providecommand{\event}{ACL2 2020} 
\usepackage{underscore}           

\title{New Rewriter Features in FGL}
\author{Sol Swords
\institute{Centaur Technology, Inc.}
\email{sswords@centtech.com}
}
\def\titlerunning{New Rewriter Features in FGL}
\def\authorrunning{S. Swords}

\usepackage{amsmath}
\usepackage{listings}
\usepackage{cite}
\usepackage{xstring}
\usepackage{breakurl}

\newcommand{\va}[1]{\ensuremath{\operatorname{\mathit{#1}}}}

\newcommand{\Code}[1]{\texttt{#1}}
\newcommand{\Codeh}[1]{\texttt{\StrSubstitute{#1}{-}{-\allowbreak{}}}}

\newcommand{\plainev}{\operatorname{\textrm{Ev}}}
\usepackage[displaymath,textmath,sections,graphics,floats]{preview}
\PreviewEnvironment{multline}
\PreviewEnvironment{lstlisting}
\PreviewMacro[{{}}]{\Code}

\begin{document}
\maketitle

\begin{abstract}
  FGL is a successor to GL, a proof procedure for ACL2 that allows
  complicated finitary conjectures to be translated into efficient
  Boolean function representations and proved using SAT solvers.  A
  primary focus of FGL is to allow greater programmability using
  rewrite rules.  While the FGL rewriter is modeled on ACL2's
  rewriter, we have added several features in order to make rewrite
  rules more powerful.  A particular focus is to make it more
  convenient for rewrite rules to use information from the syntactic
  domain, allowing them to replace built-in primitives and meta rules
  in many cases.  Since it is easier to write, maintain, and prove the
  soundness of rewrite rules than to do the same for rules programmed
  at the syntactic level, these features help make it feasible for
  users to precisely program the behavior or the rewriter.  We
  describe the new features that FGL's rewriter implements, discuss
  the solutions to some technical problems that we encountered in
  their implementation, and assess the feasibility of adding these
  features to the ACL2 rewriter.
\end{abstract}

\lstset{
    basicstyle=\footnotesize \ttfamily
  }

\section{Introduction}

FGL is a bitblasting framework for ACL2 and a successor to
GL~\cite{11-swords-bit-blasting,17-swords-term-level}.  Its aims are
similar to those of GL: to allow finitary propositions written in
idiomatic ACL2 to be solved using Boolean reasoning techniques,
including state-of-the-art SAT solving and circuit-based
simplification.  While GL approached this goal largely by including
specialized routines for symbolically simulating a large set of
primitive functions, FGL allows more user customization of its
behavior.  FGL still allows such specialized routines in the form of
metafunctions, but it has replaced many of these with rewrite rules,
which are easier for users to create, enable, and disable than the
specialized primitive functions of GL.  FGL also supports the use of
incremental SAT to solve sequences of satisfiability queries while
preserving the lemmas learned by the SAT solver from previous queries.
A further design goal of FGL is to allow sophisticated reasoning
routines built around incremental SAT to be programmed using rewrite
rules.

With these goals in mind, FGL's rewriter includes several new features
that are not present in either the GL rewriter or ACL2's rewriter.
These new features allow rewrite rules to program the rewriter more
accurately and efficiently without resorting to the use of complicated
meta rules.  We are assured of the soundness of these techniques because we
have proved FGL's correctness as a verified clause
processor~\cite{KAUFMANN20093}.  The full FGL sources, including its
soundness proof, are available in the ACL2 community
books~\cite{fgl-github}.  This paper describes these new rewriter
features, comparing them with existing features of the ACL2 rewriter.

We begin
by reviewing conditional rewriting as implemented in ACL2 in Section
\ref{secBasicRewriting}.  
In Section~\ref{fgl features sec} we describe the new rewriter
features introduced in FGL.  In Section~\ref{technical problems sec}
we discuss two technical problems that we encountered in adding these
features, and their solutions.  In Section~\ref{porting sec} we assess
the practical feasibility of adding these features to the ACL2
rewriter.


\section{Basic Rewriting}
\label{secBasicRewriting}

ACL2, GL, and FGL each use an inside-out rewriter that takes as input
a term and a subsitution alist.  ACL2's rewriter returns a new term,
whereas GL's and FGL's return a \textit{symbolic object} of a hybrid
format that incorporates both termlike constructs (variables and
function calls) and references to Boolean function objects.  We use
\textit{result objects} to refer generically to terms for the ACL2
rewriter and symbolic objects for GL/FGL.  The input substitution
alist $\sigma$ maps variables to result objects that are
considered not to need further rewriting.  Each of the rewriters
operates basically as follows, eliding many important details:
\begin{itemize}
\item If the input term is a quote, then return the quotation of its value as a result object.
\item If a variable, return its binding from the substitution alist.
\item If a lambda application, recursively rewrite the actuals, then create a new
  substitution by pairing the formals with the results from rewriting
  the actuals and recursively rewrite the body with that substitution.
\item Otherwise, a function call: recursively rewrite the arguments
  and then create the result object representing the
  call of the function on the rewritten arguments.  Apply
  rewrite rules to this object; if any rule succeeds, return its
  result, otherwise return the function call object.
\end{itemize}

The correctness contract of such a rewriter is
essentially the following equation:
\[
  \plainev{}(\va{in} \backslash \sigma, \va{env}) \equiv \plainev{}(\va{out}, \va{env})
\]
Here $\plainev{}$ is an evaluator for result objects%
---terms for the ACL2 rewriter, symbolic objects for the GL/FGL
rewriters.  The notation $x \backslash \sigma$ denotes applying
substitution $\sigma$, a mapping from variables to result objects, to
a term $x$, producing a result object.  The equivalence relation
$\equiv$ is an auxiliary input to each rewriter and is modified
according to congruence rules as the rewriters recur over terms.

In the rest of this paper we'll sometimes abuse notation by eliding
the evaluation operator and environment in equations like the one
above.  That is, if we say $x \equiv y$, where contextually $x$ and
$y$ are either both result objects or both terms, we really mean for all $\va{env}$,
$\plainev{}(x, \va{env}) \equiv \plainev{}(y, \va{env})$, where $\plainev{}$ is a result object evaluator or term evaluator as appropriate.  In this
notation we can state our rewriter correctness contract as simply
\[
 \va{in} \backslash \sigma \equiv \va{out}.
\]

The rewrite rules applied in the process described above are justified
by theorems of the form
\[
 \va{hyps} \Rightarrow \va{lhs} \equiv_R \va{rhs}
\]
where $\va{hyps}$, $\va{lhs}$, and $\va{rhs}$ are terms and
$\equiv_R$ is some equivalence relation.  Applying such a rule to a
result object $x$ is done using the following steps:
\begin{itemize}

\item Check that the equivalence $\equiv_R$ of the rule is a
  refinement of the current equivalence context $\equiv$ of the
  rewriter; that is, $a \equiv_R b$ implies $a \equiv b$.

\item Try to find a substitution $\sigma$ for which
  $x = \va{lhs} \backslash \sigma$.  This is done using a unification
  algorithm such as ACL2's \texttt{one-way-unify}.

\item Check that the hypotheses can be proved for the substitution
  $\sigma$ under the current set of assumptions.  Usually this check
  is done by backchaining, that is, recursively rewriting each of the
  hypotheses under the substitution.

\item Rewrite \va{rhs} under substitution $\sigma$ and equivalence relation
  $\equiv$ to obtain a result \va{res} for which
  $\va{res} \equiv \va{rhs} \backslash \sigma$, and return \va{res}
  as the replacement for $x$.
\end{itemize}

  
This last step is justified by the fact that, if all the previous
steps were successful, then instantiation of the rewrite rule theorem
by $\sigma$ produces $x \equiv_R \va{rhs} \backslash \sigma$, and
since $\va{rhs} \backslash \sigma \equiv \va{res}$, therefore
$x \equiv \va{res}$.



Note that the substitution $\sigma$ is
simply derived from unifying the left-hand side of the rule with the
target term.  However, this will only bind variables that appear in $\va{lhs}$.
  For any other variables, the substitution could
be extended with any assignment as long as it makes the hypotheses
true.  ACL2 takes advantage of this with its \texttt{bind-free}
feature, which FGL replicates.  Furthermore, FGL's most powerful
rewriting features are extensions of this capability, allowing free
variables to be bound in more contexts and with more flexible
semantics.

\textit{Meta rules}~\cite{05-hunt-meta} are another
form of rule that can be used for rewriting in ACL2.  A meta rule names a
function, called a \textit{metafunction}, that can be used to
syntactically transform a term to another equivalent term, and
optionally a \textit{hypothesis metafunction} that generates
hypotheses that must be relieved in order to apply that metafunction.
FGL supports meta rules as well, but its metafunctions are slightly
different in that instead of returning only a new term, they return a
term and a substitution.  FGL's meta rules do not yet support
hypothesis metafunctions.

\section{FGL Rewriter Features}
\label{fgl features sec}
We first introduce two relatively simple features of the FGL rewriter,
and then discuss a more complicated and powerful feature involving the
binding of free variables.

\subsection{\texttt{Unequiv} context}
\label{unequiv sec}
FGL's rewriter supports congruence-based rewriting much like ACL2's rewriter,
using congruence rules to determine the equivalence relations that
must be maintained on subterms when recurring through the term to be
rewritten.  (FGL only supports simple congruences, not patterned
congruences~\cite{kaufmann-equivalence-2014}.)  A simple extension to
congruence-based reasoning is special recognition of the trivial
equivalence relation under which all objects are equivalent, which we
call $\Codeh{unequiv}$\footnote{ We chose this name as both an
  abbreviation for ``universal equivalence'' and because if two
  objects are said to be \textit{unequiv}, we think it suggests that
  they're not necessarily equivalent, but not necessarily
  inequivalent either.}.  This equivalence relation is special because
when rewriting any function or lambda call under $\Codeh{unequiv}$, it
is sound to also rewrite the arguments under $\Codeh{unequiv}$---this
is also true of \Codeh{equal}, but not true of other equivalence
relations.  There are only two rewriter changes necessary to support
$\Codeh{unequiv}$: to automatically propagate the $\Codeh{unequiv}$
context into function and lambda arguments, and to recognize that all
equivalence relations are refinements of $\Codeh{unequiv}$.

Under an $\Codeh{unequiv}$ context, it is permissible to replace any
term with any result.  A rewrite rule with $\Codeh{unequiv}$ as its
equivalence relation effectively has no proof obligation, and as such can
be used to program arbitrary routines into the FGL rewriter.  FGL also
allows certain special features under an $\Codeh{unequiv}$ context that
would be unsound otherwise:
\begin{itemize}
\item $\Codeh{syntax-interp}$ evaluates its argument term as in ACL2's
  \Codeh{syntaxp} or \Codeh{bind-free}, that is, under the assignment of
  each variable to the result object bound to it in the current substitution.
\item $\Codeh{fgl-interp-obj}$ rewrites its argument term as usual,
  then if its result is the quotation of a term, it calls the rewriter
  on that term.
\item $\Codeh{assume}$ rewrites its first argument, then assumes that
  result to be true while rewriting its second argument.
\end{itemize}

Users can prove congruence rules that induce an $\Codeh{unequiv}$
context on any function argument that is irrelevant to the value of
that function.  For example, we define
$\Codeh{(fgl-prog2 x y)} = \Codeh{y}$, and provide a congruence rule
that induces an $\Codeh{unequiv}$ context on the first argument of
\Codeh{fgl-prog2}.  This can be used to perform extralogical analyses
and print results as a side effect.  We also provide a utility
$\Codeh{(bind-var x y)}$ which must occur with $x$ a free variable;
this rewrites $y$ under an $\Codeh{unequiv}$ context and binds $x$ to
its result.  This is often used in combination with
$\Codeh{syntax-interp}$ to obtain $\Codeh{bind-free}$-like functionality
that can be used in the midst of a rule's right-hand
side as well as in the hypotheses.

\subsection{\texttt{abort-rewrite}}

It is always sound to decide not to apply a rule.  FGL supports a
special identity function $\Codeh{abort-rewrite}$ which causes the
rewriter to abort the current rule attempt whenever it is encountered.
This allows rules to be programmed such that while rewriting the
right-hand side, some condition may cause the rule not to be applied
after all.  The logical definition of $\Codeh{abort-rewrite}$ is an
identity function so that authors of such rules may wrap it around
whatever term is most convenient for proving the rule correct.  The
wrapped term will not be rewritten when applying the rule; instead,
the application of the rewrite rule will be aborted.

An example of the use of \Codeh{abort-rewrite} is shown in
Listing~\ref{fgl equal def}, which shows a rewrite rule that can be
used for resolving many calls of \Code{equal}.  In this rule,
functions such as \Codeh{check-integerp} are binder functions,
discussed in Section~\ref{free variable sec}, and calls such as
\Code{(check-integerp x-intp x)} return true if \Code{x} is
syntactically known to be an integer.  This rule uses
\Codeh{abort-rewrite} in several situations where it doesn't know how
to resolve the equality of the inputs.  In such cases, application of
this rule fails, but the target object may still be rewritten by other
rules.

\begin{lstlisting}[float, label=fgl equal def, caption={Example rule using \Code{abort-rewrite} and binder functions}]
(def-fgl-rewrite fgl-equal
  (equal (equal x y)
         (cond ((check-integerp x-intp x)
                (cond ((check-integerp y-intp y)
                       (and (iff (intcar x) (intcar y))
                            (or (and (check-int-endp x-endp x)
                                     (check-int-endp y-endp y))
                                (equal (intcdr x) (intcdr y)))))
                      ((check-non-integerp y-non-intp y) nil)
                      (t (abort-rewrite (equal x y)))))
               ((check-booleanp x-boolp x)
                (cond ((check-booleanp y-boolp y)
                       (iff x y))
                      ((check-non-booleanp y-non-boolp y) nil)
                      (t (abort-rewrite (equal x y)))))
               ((check-consp x-consp x)
                (cond ((check-consp y-consp y)
                       (and (equal (car x) (car y))
                            (equal (cdr x) (cdr y))))
                      ((check-non-consp y-non-consp y) nil)
                      (t (abort-rewrite (equal x y)))))
               ((and (check-integerp y-intp y)
                     (check-non-integerp x-non-intp x)) nil)
               ((and (check-booleanp y-boolp y)
                     (check-non-booleanp x-non-boolp x)) nil)
               ((and (check-consp y-consp y)
                     (check-non-consp x-non-consp x)) nil)
               (t (abort-rewrite (equal x y))))))
\end{lstlisting}

\subsection{Free Variable Binding}
\label{free variable sec}

The choice of bindings for variables not present in the left-hand side
of a rewrite rule is a powerful tool.  Since free variables can be
bound to anything, they can be bound based on extralogical
considerations such as the syntax of the term being rewritten. This
flexibility can help give rewrite rules the power of metafunctions
without needing to program them wholly at the syntactic level.

Like the ACL2 rewriter, the FGL rewriter supports \Codeh{bind-free}
hypotheses~\cite{05-hunt-meta}.  These allow arbitrary bindings to
computed based on term syntax and added to the substitution that will
be used in applying the rule.  Both rewriters also support free
variable bindings based on equivalence hypotheses \Codeh{(equiv var
  term)}, described in the ACL2 documentation topic
\Codeh{free-variables}~\cite{acl2:doc}.  FGL also adds a new way of
binding free variables using \textit{binder rules}.  Binder rules may
be used to bind free variables in the right-hand side of rules as well
as the hypotheses.  They provide wide flexibility in the strategies
used to choose the bindings, from rewriting (as in equivalence-based
binding hypotheses) to syntactic interpretation as in
\Codeh{bind-free}.  In fact, the \Codeh{bind-var} utility described in
Section~\ref{unequiv sec} is implemented using a binder rule.
Finally, they also allow facts about how the variable will be bound to
be used in the logical justification for the rewrite rules in which
they occur.

The rule \Codeh{fgl-equal} shown in Listing~\ref{fgl equal def}
provides a basic example of the kind of rewriter programming that may
be done with binder rules.  Each of the functions prefixed
\Codeh{check-} is a binder function which effectively checks whether
the argument \Code{x} or \Code{y} satisfies some syntactic criteria;
for example, \Codeh{(check-integerp x-intp x)} returns \Code{t} if
\Code{x} is syntactically known to be an integer.  This is
accomplished by binding the free variable \Codeh{x-intp} to the result
of the syntactic check.  To do this in a rule written for the ACL2
rewriter, one would need to perform all of these free variable
bindings using \Codeh{bind-free} hypotheses, rather than as needed in
the right-hand side.  It is also known in the logic that
\Code{(check-integerp x-intp x)} implies that \Code{x} is an integer.
In contrast, \Codeh{bind-free} doesn't provide any information in the
logic about the free variables that it binds, so the result of a
syntactic check that \Code{x} was an integer would need to be paired
with a symbolic check such as \Code{(integerp x)}, resolved by further
rewriting.





The fact that \Codeh{bind-free} and the related utility \Codeh{syntaxp} give no information in
the logic about the syntactic computation performed leads
to awkward usage patterns.  Often we know that a term satisfies some
property, either by a syntax check or by construction, but we still
must add a hypothesis checking that property or we won't be able to
prove the rule.  Here are two examples from the
``rtl/rel9'' library of the ACL2 community books~\cite{acl2:doc, russinoff-book-2019}, preceded by the definition of \Code{power2p} used therein:
\begin{lstlisting}
(defund power2p (x)
  (declare (xargs ...))
  (cond ((or (not (rationalp x))
             (<= x 0))
         nil)
        ((< x 1) (power2p (* 2 x)))
        ((<= 2 x) (power2p (* 1/2 x)))
        ((equal x 1) t)
        (t nil) ;got a number in the doubly-open interval (1,2)
        ))

(defthm power2p-shift-2
  (implies (and (syntaxp (power2-syntaxp y))
                ;this should be true if the syntaxp hyp is satisfied
                (force (power2p y)))
           (equal (power2p (* x y))
                  (power2p x))))

(defthm expo-shift-general
  (implies (and (bind-free (bind-k-to-common-expt-factors x) (k))
                ...
                (force (power2p k))
                ...
                )
           (equal (expo x)
                  (+ (expo k) (expo (* (/ k) x))))))
\end{lstlisting}
In both theorems the \Code{(force (power2p \va{v}))} hypothesis is
checking something that is already known.  The syntax check
\Codeh{power2-syntaxp} provably is only true of terms whose evaluation
satisfies \Codeh{power2p}, and the binding function
\Codeh{bind-k-to-common-expt-factors} provably will only bind the
variable \Codeh{k} to a term satisfying \Codeh{power2-syntaxp}.
However, the author of these rules couldn't use these facts when
proving the theorems justifying them, so the forced \Codeh{power2p}
hypotheses were used instead.  When these rules are applied, these
redundant hypotheses must be relieved using rewriting.

Arguably, such a redundant check is a small price to pay to be able to
use syntactic checks and arbitrary free variable bindings.  In many
cases the checks can likely be optimized so that they only need to
repeat a small amount of work.  However, even the single rule
\Codeh{power2p-shift-2} above demonstrates that blowups are possible:
note that if this rule is used to prove \Codeh{power2p} of a product of
size $n$, then \Codeh{power2-syntaxp} will run $\Theta (n^2)$ times
since each top-level call recurs through the whole product. Without
the redundant hypothesis, only one linear check would be
necessary.

To help avoid the need for these redundant checks, FGL adds
\textit{binder rewrite} and \textit{binder meta} rules.  A binder rule
of either kind acts on a \textit{binder function}.  The binder
function expresses the properties that the free variable's eventual
binding must satisfy, and the binder rule dictates how the variable
will be bound.  The binder function's first argument (per our
convention) is the free variable, which the function fixes so that
it satisfies some property that may depend on the rest of the
arguments.  That is, if the first argument satisfies the desired
properties, it is passed through unchanged; otherwise, some other
value that does satisfy the properties is returned instead.  

Listing~\ref{binder fn examples} shows several examples of binder
functions without their binder rule implementations.  First, \Codeh{bind-var} simply returns the first argument unchanged, so it places no restriction on the binding.  The next,
\Codeh{syntactically-true}, could be implemented as a check that
\Codeh{x} is either a constant nonnil value or a call of some function
known not to return \Codeh{nil}---or, at the most conservative, the
implementation could always bind the variable to \Codeh{nil}.  Third, \Codeh{integer-length-bound} produces
an upper bound for the integer-length of \Codeh{x}, or \Codeh{nil} if
none can be found.  The final, and most complex,
\Codeh{split-list-by-membership} returns two lists which
must satisfy two criteria: the two lists together must be
set-equivalent to the input list \Codeh{x}, and elements of the first
list must be members of the second input list \Codeh{y}.  Crucially,
the use of a free variable allows the split not to be a function of
\Codeh{x} and \Codeh{y}.  The implementing binder rule could use various
levels of effort to check whether elements of \Codeh{x} are verifiable members of
\Codeh{y} so that it can put them in the first list.

\begin{lstlisting}[float, label=binder fn examples, caption={Binder function examples}]
;; No restrictions
(defun bind-var (var x)
  var)

;; True only if x is true
(defun syntactically-true (var x)
  (and x var))

;; Upper bound for (integer-length x) if nonnil
(defun integer-length-bound (var x)
  (and (integerp var)
       (<= (integer-length x) var)
       var))

;; All elements of x are in either the first or second return value,
;; and all elements of the first return value are in y
(defun-nx split-list-by-membership (var x y)
   (mv-let (part1 part2) var
       (if (and (set-equiv (append part1 part2) x)
                (subsetp-equal part1 y))
           (mv part1 part2)
         (mv nil x))))
\end{lstlisting}

Binder rules resolve calls of binder functions by producing a term
that can be consistently used simultaneously as the new binding for
the free variable and the replacement for the binder function call.
A binder rewrite rule is justified by a theorem of the form
\[
\va{hyps} \wedge ( \va{var} \equiv_H \va{form} ) \Rightarrow f(\va{var}, \va{args}) \equiv_C \va{var}
\]
where \va{hyps} and \va{form} are terms, \va{args} is an argument list of zero or
more terms, \va{var} is a variable not present in any of those terms,
and $f$ is the target binder function.
Applying such a rule to a call $f(v, \va{args}')$, with $v$ a free variable, is done using the following
steps:
\begin{itemize}
\item Check that the equivalence $\equiv_C$ of the rule is a
  refinement of the rewriter's current equivalence context $\equiv$.

\item Try to find a substitution $\sigma$ for which
  $\va{args}' = \va{args} \backslash \sigma$, using one-way
  unification.

\item Relieve the hypotheses under substitution $\sigma$ by backchaining.

\item Rewrite \va{form} under substitution $\sigma$ and equivalence
  $\equiv_H$ to obtain a result \va{res} for which
  $\va{res} \equiv_H \va{form} \backslash \sigma$.  Add $\va{res}$ as
  the binding for $v$ and return it as the replacement for the call of
  $f$.

\end{itemize}

The last step is justified because, if all the previous steps were
successful, then instantiation of the binder rule theorem by
$\{ \va{var} \leftarrow v \} \cup \sigma$ produces
\[
 ( v \equiv_H \va{form} \backslash \sigma ) \Rightarrow f(v, \va{args}') \equiv_C v
\]
(using the assumption that $\va{var}$ does not appear in \va{hyps},
\va{form}, or \va{args} as noted above).  Since $v$ is a free variable
we may bind it however we want; binding it to \va{res} ensures that
the antecedent of the above implication holds, so that we may replace
the call of $f$ with \va{res} as well.

For the binder function examples above, we'll now describe binder rules
that implement the intended checks.  The binder rule for
\Codeh{bind-var} is shown in Listing~\ref{bind var impl}; also
important in its implementation is a congruence rule which allows
rewriting its second argument under \Codeh{unequiv}.  When a
\Codeh{bind-var} form is encountered with its first argument a free
variable, the second argument is first rewritten under \Codeh{unequiv}
due to the congruence rule, and the variable is then bound to that
result due to the binder rule.

\begin{lstlisting}[float=p, label=bind var impl, caption={Bind-var implementation}]
(defcong unequiv equal (bind-var var x) 2)
  (add-fgl-congruence unequiv-implies-equal-bind-var-2)
  
(def-fgl-brewrite bind-var-binder-rule
   (implies (equal var x)
            (equal (bind-var var x) var)))
\end{lstlisting}

We can implement \Codeh{syntactically-true} using
a pair of binder rewrite rules, as listed in
Listing~\ref{syntactically true binder}.  The second
rule will be tried before the first, assuming they are submitted in
that order.  The second handles the case where \Codeh{x} is known to be
true, assigning \Codeh{var} the value \Codeh{t}.  The first rule applies
if the second fails, in which case \Codeh{var} is assigned \Codeh{nil}.
\begin{lstlisting}[float=p, label=syntactically true binder, caption={Syntactically-true implementation}]
(def-fgl-brewrite syntactically-true-binder-rewrite-false
  (implies (equal var nil)
           (equal (syntactically-true var x) var)))
  
(def-fgl-brewrite syntactically-true-binder-rewrite-true
  (implies (and x (equal var t))
           (equal (syntactically-true var x) var)))
\end{lstlisting}

For \Codeh{integer-length-bound}, we could use several rewrite rules to
handle different cases like we did with \Codeh{syntactically-true}.
Instead, we show in Listing~\ref{integer length bound rule} an
implementation that uses one rewrite rule, programmed in a more
explicit style.  The rule checks three possible
cases, based on whether \Codeh{x} is syntactically a symbolic integer,
a concrete (quoted) value, or neither.  If it is neither, then it
produces \Codeh{nil} as the new binding.  If it is a concrete value,
then it returns its exact integer length.  For symbolic integers, it
checks whether \Codeh{(int-endp x)} is known to be true---that is,
\Codeh{x} is a non-integer, 0, or -1.  If so, then its integer-length
is 0.  Otherwise, we bound the integer-length of the \Codeh{logcdr}
(right-shift by 1) of \Codeh{x}, and add 1 to the result if it is
non-nil.

\begin{lstlisting}[float=p, label=integer length bound rule, caption={Integer-length-bound implementation}]
(def-fgl-brewrite integer-length-bound-binder-rw
  (implies
    (equal var (cond ((bind-var symbolic (syntax-interp
                                          (fgl-object-case x :g-integer)))
                      (if (syntactically-true known-int-endp (int-endp x))
                          0
                        (let ((rest-bound (integer-length-bound
                                           rest-bound (logcdr x))))
                          (and rest-bound (+ 1 rest-bound)))))
                     ((bind-var concrete (syntax-interp
                                          (fgl-object-case x :g-concrete)))
                      (integer-length x))
                     (t nil)))
    (equal (integer-length-bound var x) var)))
\end{lstlisting}
\begin{lstlisting}[float=p, label=split list rule, caption={split-list-by-membership implementation}]
(def-fgl-brewrite split-list-by-membership-binder-rule
  (implies (equal var (if (syntactically-true known-consp (consp x))
                          (mv-let (rest1 rest2)
                            (split-list-by-membership rest-call (cdr x) y)
                            (if (syntactically-true known-member
                                                    (member-equal (car x) y))
                                (mv (cons (car x) rest1) rest2)
                              (mv rest1 (cons (car x) rest2))))
                        (mv nil x)))
           (equal (split-list-by-membership var x y) var)))
\end{lstlisting}

This rule could potentially be improved by using a more specialized
test for \Codeh{int-endp}; as is, a full expression for \Codeh{int-endp}
must be computed for each tail of \Codeh{x}.  Another approach to
implementing \Codeh{integer-length-bound} is to use a binder meta rule.
A binder meta rule allows a certain metafunction to be used to
generate a binding for a binder function call.  The metafunction takes
the binder function name $f$ and argument objects \va{args} as input
and returns \va{form} and $\sigma$.  The theorem justifying a meta
rule says that evaluation of these terms always satisfies the binder
rewrite rule formula; that is, $\va{form} \backslash \sigma$ is always
an object that is preserved by the application of $f$ with the given
arguments.  For \Codeh{integer-length-bound}, the metafunction could
count the bits in the symbolic integer representation directly, rather
than doing it by iterative rewriting.

Finally, an implementation of
\Codeh{split-list-by-membership} is shown in
Listing~\ref{split list rule}.  This is conceptually similar to the
\Codeh{integer-length-bound} rule, recurring down a list and
conservatively crafting a result that satisfies the restriction
imposed by the binder function; we include it to illustrate the
variety of types of constraints that can be handled by these rules.

\section{Technical Problems and Solutions}
\label{technical problems sec}
The free variable binding features of FGL led to two technical
problems.  We describe those problems and their solutions in this
section.

\subsection{Free Variable Binding with Lambdas}
\label{free var lambdas sec}
One of the design goals for the FGL rewriter was to allow free
variables to be bound anywhere in a rewrite rule.  The \Codeh{bind-var}
and binder rule features may bind variables in the midst of rewriting
a term such as a hypothesis or the right-hand side.  Logically, we
just require that the variable was not previously bound, and that the
binding site of the variable was its first use.  However, ACL2's
handling of \Codeh{let}/\Codeh{lambda} expressions poses a problem.  A
well-formed lambda term in ACL2 must bind all the free variables of
its body.  For example, the translation of \Codeh{(let ((b (b-expr)))
  (f a b))} will be a call of a lambda that has both \Codeh{a} and
\Codeh{b} as formals, such as \Codeh{((lambda (a b) (f a b)) a
  (b-expr))}.  What happens, then, if we want to bind a free variable
inside a lambda body?  For example, suppose \Codeh{f} in the \Codeh{let}
expression above was a binder function, and \Codeh{a} a free variable.
Unfortunately, \Codeh{a} appears in the lambda arguments, which means
the rewriter would first encounter it as an argument to the lambda call,
not at its binding site.

To work around this problem, we use a different strategy for handling
lambdas than the ACL2 rewriter or the usual form of ACL2 evaluator.
The usual way is to first process (rewrite or evaluate) the actuals of
the lambda call under the current variable bindings (call them
$\sigma_0$), then pair the lambda formals with the results of
processing the actuals to create a new set of variable bindings
$\sigma_1$, and use these bindings to process the body of the lambda.
Instead, before we begin rewriting the actuals, we strip out any
self-pairings from the formal/actual pairs, such as the pairing of
\Codeh{a} with itself in the example above.  This allows FGL's
rewriter to avoid encountering free variables such as \Codeh{a} before
their intended binding sites.  We rewrite the remaining actuals under
$\sigma_0$ and create a variable binding alist $\sigma_2$ by pairing
the remaining formals with the results.  However, instead of using
$\sigma_2$ by itself when processing the lambda body, we append it to
the existing variable bindings and use the combined bindings
$\sigma_2::\sigma_0$, where the bindings of $\sigma_2$ shadow those of
$\sigma_0$.  The critical fact showing the semantic equivalence of
these two strategies is that each variable present in the lambda body
has the same binding in $\sigma_1$ as in $\sigma_2::\sigma_0$.  In
particular, a variable that was formerly self-paired in the lambda
call will not be present in $\sigma_2$, so its binding in
$\sigma_2::\sigma_0$ is its binding from $\sigma_0$.  Its binding in
$\sigma_1$ is derived by evaluating or rewriting the variable itself
under $\sigma_0$.  Since evaluating or rewriting a variable is done by
looking it up in the bindings, these are equivalent.

A further subtlety is that when binding a free variable, its binding
can't be local to a lambda, but must be set in the substitution
$\sigma$ of the current rewrite rule application.  Therefore the FGL rewriter
actually tracks two sets of bindings:
$\sigma_u$, the unifying substitution plus any free variables that have
been bound so far, and $\sigma_\lambda$, the combined bindings from
the current nesting of lambdas.  A variable's binding is found by
looking it up first in $\sigma_\lambda$, then in $\sigma_u$ if not
found.

\subsection{Inductive Correctness with Free Variable Bindings}

Recall that we stated the correctness contract of a rewriter, in
Section ~\ref{secBasicRewriting}:
\[
  \plainev{}(\va{in} \backslash \sigma, \va{env}) \equiv \plainev{}(\va{out}, \va{env})
\]
In FGL,  the substitution $\sigma$ is both an input to and output from
the rewriter.  For the current discussion we will ignore the fact
mentioned previously that the full substitution consists of two parts,
$\sigma_u$ and $\sigma_\lambda$; here, $\sigma$ represents both.
In each call of the rewriter, the current
substitution $\sigma_i$ is passed in, and a modified substitution $\sigma_o$, potentially
with some new free variables bound, is returned.  The correctness
statement we want has to do with the resulting substitution
$\sigma_o$:
\[
  \plainev{}(\va{in} \backslash \sigma_o, \va{env}) \equiv \plainev{}(\va{out}, \va{env})
\]
However, as stated this property is not inductive.  Consider rewriting
a function call of two arguments $f(a, b)$; we'll try and show that
the rewriter's treatment of this term is correct assuming inductively
that it correctly rewrites $a$ and $b$.  Suppose we start with
substitution $\sigma_i$.  First we rewrite $a$ with this substitution,
producing output $a'$ and substitution $\sigma_a$.  Then then rewrite
$b$ with substitution $\sigma_a$, producing output $b'$ and
subsitution $\sigma_b$.  Then for simplicity suppose we have no rewrite rules about $f$, so we just return $f(a', b')$ and substitution $\sigma_b$.  The facts we may assume inductively are:
\[
  \plainev{}(a \backslash \sigma_a, \va{env}) \equiv \plainev{}(a', \va{env})
\]
\[
  \plainev{}(b \backslash \sigma_b, \va{env}) \equiv \plainev{}(b', \va{env})
\]
We want to prove:
\[
  \plainev{}(f(a, b) \backslash \sigma_b, \va{env}) \equiv \plainev{}(f(a',b'), \va{env})
\]
Basic facts about evaluation and substitution reduce this to:
\[
  f\left(\plainev{}\left(a \backslash \sigma_b, \va{env}\right), \plainev{}\left(b \backslash \sigma_b, \va{env}\right)\right) \equiv
f\left(\plainev{}\left(a', \va{env}\right), \plainev{}\left(b', \va{env}\right)\right)
\]
Notice we have an assumption about
$\plainev{}(a \backslash \sigma_a, \va{env})$ where we need a fact
about $\plainev{}\left(a \backslash \sigma_b, \va{env}\right)$.  With
only these induction hypotheses, we are stuck.  However, a basic
property of the rewriter is that $\sigma_o$ is an extension of
$\sigma_i$---for any variable bound in $\sigma_i$, it must be bound to
the same value in $\sigma_o$.  We'll denote this using set notation as
$\sigma_i \subseteq \sigma_o$.  Intuitively, $\sigma_a$ should bind
all the variables that are needed to evaluate $a$, so that the new
variables bound in $\sigma_b$ don't affect the result.  Therefore, the
inductive assumption we need is that the evaluation equivalence holds
for any extension to the resulting substitution:
\[
  \forall \sigma_+ : \sigma_o \subseteq \sigma_+ \Rightarrow \plainev{}(\va{in} \backslash \sigma_+, \va{env}) \equiv \plainev{}(\va{out}, \va{env}).
\]
Trying our proof again, we have inductive assumptions:
\[
  \forall \sigma_+ : \sigma_a \subseteq \sigma_+ \Rightarrow \plainev{}(a \backslash \sigma_+, \va{env}) \equiv \plainev{}(a', \va{env})
\]
\[
  \forall \sigma_+ : \sigma_b \subseteq \sigma_+ \Rightarrow \plainev{}(b \backslash \sigma_+, \va{env}) \equiv \plainev{}(b', \va{env})
\]
and we want to prove
\[
  \forall \sigma_+ : \sigma_b \subseteq \sigma_+ \Rightarrow \plainev{}(f(a, b) \backslash \sigma_+, \va{env}) \equiv \plainev{}(f(a',b'), \va{env})
\]
or equivalently
\[
  \forall \sigma_+ : \sigma_b \subseteq \sigma_+ \Rightarrow 
  f\left(\plainev{}\left(a \backslash \sigma_+, \va{env}\right), \plainev{}\left(b \backslash \sigma_+, \va{env}\right)\right) \equiv
f\left(\plainev{}\left(a', \va{env}\right), \plainev{}\left(b', \va{env}\right)\right).
\]
Fortunately, we can apply the transitivity of substitution extension
to conclude $\sigma_a \subseteq \sigma_+$ so that this time we may
apply the induction hypothesis about $a$ as well as the one about $b$.
Of course, the full proof that the FGL rewriter is correct is beyond
the scope of this paper, but may be examined in the ACL2 community
book ``centaur/fgl/interp.lisp.''

\section{Porting to the ACL2 Rewriter}
\label{porting sec}
If these features of the FGL rewriter are useful, it raises the
question of whether they could be ported to the ACL2 rewriter.  In
this section we try to assess how difficult it would be to add these
features, whether they conflict with any other features, etc.  These
assessments are based on examination of the relevant ACL2 code, and
not on any attempt to implement them; therefore, there might be
impediments that we didn't foresee.

\subsection{\texttt{Unequiv} context}

Support for \Codeh{unequiv} would the easiest of these features to add
to the ACL2 rewriter.  We believe the following changes would be the
only ones necessary:
\begin{itemize}
\item Set \Codeh{unequiv} as the equivalence context for all arguments
  whenever a function or lambda call occurs in \Codeh{unequiv} context.
  This could be done by adding a special case to the function \Codeh{geneqv-lst}, which
  computes the equivalence contexts for rewriting a function's
  arguments given the equivalence context in which the function is
  being rewritten.
\item Recognize that any other equivalence relation is a refinement of
  \Codeh{unequiv}.  This could be done by adding a special case to the
  function \Codeh{geneqv-refinementp}, which determines whether an
  equivalence is a refinement of the current equivalence context.
\end{itemize}
These changes would suffice for allowing the rewriter to use
\Codeh{unequiv}-based rewrite rules in logically irrelevant contexts.
Additional features that are allowed only under \Codeh{unequiv}
context, such as \Codeh{syntax-interp}, could be considered separately.

\subsection{\texttt{abort-rewrite}}

In order to implement \Codeh{abort-rewrite}, the call of \Codeh{rewrite}
in the function \Codeh{rewrite-with-lemma} would need to return a flag
saying that the rule application attempt failed.  This would affect
most of the mutually recursive clique implementing the rewriter, though not in a very deep
way. Rewriter functions \Codeh{rewrite}, \Codeh{rewrite-if},
\Codeh{rewrite-args}, \Codeh{rewrite-with-lemmas}, and
\Codeh{rewrite-fncall} would all need to return the abort flag, and
callers of those functions in \Codeh{rewrite-equal},
\Codeh{relieve-hyp}, \Codeh{rewrite-with-lemma},
\Codeh{rewrite-linear-term}, and \Codeh{multiply-alists2} would need to
deal with the possibility of an abort.  Rather than inserting tests in
order to exit early after calls that produce an abort, it might be
cleaner to pass the abort flag along through the rewriter, checking
for it at a convenient point such as the entry to \Codeh{rewrite}.

As an additional consideration, it would be unfortunate to abort a
proof completely due to the presence of \Codeh{abort-rewrite} in the
statement of the conjecture to be proved.  This could be dealt with by
adding an input flag to the rewriter (perhaps part of the \Codeh{rcnst}
structure) that says whether \Codeh{abort-rewrite} calls are to be
respected, or by simply stripping out \Codeh{abort-rewrite} calls from
the conjecture before attempting its proof.

\subsection{Free Variable Binding}

At first glance adding the ability to bind free variables anywhere
would seem not to touch any more code than \Codeh{abort-rewrite}.
Essentially it would require passing the unifying substitution out of
all the same functions that would need to return the abort flag.  One
could even get clever and combine the unifying substitution with the
abort flag, since the substitution is irrelevant if the rule
application is to be aborted.

However, there is another logical issue to work out.  When relieving
hypotheses, the ACL2 rewriter is careful to ensure that variables
aren't used before they're bound in the unifying substitution.  But
otherwise, it assumes that a variable not bound in the
substitution is implicitly bound to itself.  It uses the following form
to look up variables in the substitution, where \Codeh{term} is the
variable:
\begin{lstlisting}
(let ((temp (assoc-eq term alist)))
  (cond (temp (cdr temp))
        (t term)))
\end{lstlisting}
In fact, in several places the ACL2 rewriter uses alist \Codeh{nil} to
signify that all variables in the term to be rewritten are bound to
themselves.  If we allowed free variables to be bound arbitrarily
within terms being rewritten, we'd need to know that they hadn't been
previously assumed to be bound to themselves.  Otherwise, we could
prove \Codeh{nil} by applying something like the following rule:
\begin{lstlisting}
(equal (always-true)
       (equal a (bind-var a (syntax-interp '(not a)))))
\end{lstlisting}
Additionally, in order to be able to bind variables within \Codeh{let}
or \Codeh{lambda} expressions, the ACL2 rewriter would need to adopt a
similar style of variable binding as we discussed in
Subsection~\ref{free var lambdas sec}, splitting the variable
assignment into the unifying substitution and local lambda bindings
and filtering out self-bindings from lambdas.

\begin{lstlisting}[float, label=bind split list example, caption={Binder hypothesis rule concept}]
(defun-nx bind-split-list-by-membership (free-var x y)
  (mv-let (part1 part2) free-var
    (and (set-equiv (append part1 part2) x)
         (subsetp-equal part1 y))))

(defthm bind-split-list-by-membership-default
   (implies (equal free-var (mv nil x))
            (bind-split-list-by-membership free-var x y))
   :rule-classes :binder-hyp)

(defthm bind-split-list-by-membership-nonmember
  (implies (and (consp x)
                (bind-split-list-by-membership rest-split (cdr x) y)
                (equal free-var (mv-let (rest1 rest2) rest-split
                                   (mv rest1 (cons (car x) rest2)))))
           (bind-split-list-by-membership free-var x y))
   :rule-classes :binder-hyp)

(defthm bind-split-list-by-membership-member
  (implies (and (consp x)
                (member (car x) y)
                (bind-split-list-by-membership rest-split (cdr x) y)
                (equal free-var (mv-let (rest1 rest2) rest-split
                                   (mv (cons (car x) rest1) rest2))))
           (bind-split-list-by-membership free-var x y))
   :rule-classes :binder-hyp)
\end{lstlisting}

If it isn't feasible to add the capability of binding free variables
within a term, something akin to binder rules could still be pursued
for use on top-level hypotheses.  Instead of binder functions that
fix a free variable so that it complies with some property, we could
use binder hypothesis functions that simply describe the properties
satisfied by a free variable binding, and binder hypothesis rules that
say how to bind those free variables in such a way that the binder
hypothesis function is true.  An example based on the
\Codeh{split-list-by-membership} binder described in Subection~\ref{free
  variable sec} is shown in Listing~\ref{bind split list example}.

Adding this capability would require many changes as well. Presumably,
binder hypothesis rules would need to have their own rule class, with
support for processing appropriate forms of \Codeh{defthm} as well as
support for applying the rules in the rewriter within
\Codeh{relieve-hyp}.  Binder hypothesis meta rules would likely be an
incremental addition on top of this.

\section{Conclusion}
FGL's rewriter adds new features that allow a style of metaprogramming
via rewrite rules that cannot practically be done with the ACL2
rewriter's current feature set.  The new free variable binding
capabilities allow syntactic information to be used in directing the
rewriter without losing (or needing to re-verify) the semantic
information that the syntactic properties imply.  Rewrite rules using
these capabilities provide an alternative to complicated metafunctions
that must be proved correct relative to an evaluator.  We hope to show
in future work that it also provides a platform on which significant
proof and analysis routines can be built quickly and easily, using
powerful automatic tools such as incremental SAT and Boolean circuit
simplifiers.

\section*{Acknowledgements}
I would like to thank Matt Kaufmann for maintaining ACL2, and more
specifically for working out the removal of a restriction involving
the use of attachments in metafunctions which has been crucial to the
development of this work.  I'd also like to thank my colleagues at
Centaur Technology, Shilpi Goel, Anna Slobodova, and Rob Sumners, for
their help and encouragement in developing FGL.

\bibliographystyle{eptcs}
\bibliography{../bib.bib}
\end{document}